\documentclass[jou,floatsintext]{apa7}

\usepackage{subcaption}
\usepackage{tipa}
\usepackage{amsmath}
\usepackage{setspace}

\usepackage[natbib,backend=biber,style=apa]{biblatex}
\usepackage{lineno,hyperref}

\title{Gradient boundaries through confidence intervals for forced alignment estimates using model ensembles}
\shorttitle{Gradient boundaries for forced alignment}
\leftheader{Kelley}

\authorsnames[{1}]{Matthew C. Kelley}
\authorsaffiliations{{Linguistics Program, Department of English, George Mason University, Fairfax, Virginia, 22030, USA}}

\authornote{
\addORCIDlink{Matthew C. Kelley}{0000-0002-7218-5599}

Correspondence should be addressed to Matthew C. Kelley. Email: mkelle21@gmu.edu.
}


\begin{document}

\abstract{    
    Forced alignment is a common tool to align audio with orthographic and phonetic transcriptions. Most forced alignment tools provide only point-estimates of boundaries. The present project introduces a method of producing gradient boundaries by deriving confidence intervals using neural network ensembles. Ten different segment classifier neural networks were previously trained, and the alignment process is repeated with each classifier. The ensemble is then used to place the point-estimate of a boundary at the median of the boundaries in the ensemble, and the gradient range is placed using a 97.85\% confidence interval around the median constructed using order statistics. Gradient boundaries are taken here as a more realistic representation of how segments transition into each other. Moreover, the range indicates the model uncertainty in the boundary placement, facilitating tasks like finding boundaries that should be reviewed. As a bonus, on the Buckeye and TIMIT corpora, the ensemble boundaries show a slight overall improvement over using just a single model. The gradient boundaries can be emitted during alignment as JSON files and a main table for programmatic and statistical analysis. For familiarity, they are also output as Praat TextGrids using a point tier to represent the edges of the boundary regions.
}

\maketitle

\setcounter{secnumdepth}{3}

\section{Introduction}

Forced alignment is a tool used to annotate where words and/or segments occur in an audio signal. Forced aligners take in an audio file, an orthographic transcription, and a so-called ``grapheme to phoneme'' dictionary. The product is a time-alignment for each segment in the signal, and this alignment is often used to produce a TextGrid with segment and word boundaries usable in Praat \citep{boersma2001praat}, though other formats may also be produced. Such tools are still widely used across research in phonetics and speech science \citep{mahrPerformanceForcedAlignmentAlgorithms2021,yuanUsingForcedAlignment2023} and were historically part of how automatic speech recognition systems were developed with hidden Markov models. These systems only provide a point estimate of the boundaries between segments and do not, traditionally, yield a gradient boundary region or quantify the uncertainty around the boundary placement. The present paper introduces a method for producing gradient boundary regions for forced alignment. These regions are achieved by deriving confidence intervals through using ensemble methods.

Ensemble methods---such as model averaging---are not unheard of in machine learning \citep{breimanBaggingPredictors1996,carneyConfidencePredictionIntervals1999}. They are prominent enough that competition-winning models have used such techniques to improve their performance \citep{korenBellKorSolutionNetflix2009,krizhevskyImageNetClassificationDeep2012}, and they are basic enough to be covered in textbooks \citep{goodfellowDeepLearning2016}. Despite their potential to offer uncertainty estimates of model predictions, ensemble methods do not yet seem to have been used in creating end-user forced alignment systems. However, they have recently been used in other areas of speech science, such as for modeling vocal physiology with included confidence intervals \citep{zhangAmbulatoryMonitoringVocal2025}.

A system that produces estimates of boundary regions in forced alignment models has several potential benefits. One of these is that ambiguous segment boundaries can be more reasonably handled than when only point estimates are produced. Another is that researchers can more easily determine where the alignment model has made outlying predictions and caused significant errors in alignment.

The present work demonstrates the calculation of gradient boundaries via confidence intervals by extending the Mason-Alberta Phonetic Segmenter (MAPS) system \citep{kelleyMasonAlbertaPhoneticSegmenter2024}. MAPS is structured as a deep recurrent neural network that was trained as a phone recognizer. The network accepts mel frequency cepstral coefficients (MFCCs) and outputs the probability of each phone given the MFCCs. The alignment boundaries are derived from the probabilities calculated for an utterance using the \texttt{Decode} algorithm \citeauthor{kelleyMasonAlbertaPhoneticSegmenter2024} specified.

This innovation in segment boundary representation addresses a particular shortcoming in the field that our annotations do not fully reflect the way we view speech. Despite being meaningfully formalized in, for example, articulatory phonology \citet{browmanArticulatoryPhonologyOverview1992}, the overlapping nature of speech has not entered into the data formats that we commonly use in speech data annotation. Rather, we use annotation schemes, like TextGrids in Praat \citep{boersma2001praat}, which enforce no overlap between annotations. In part, this is a simplification to make it easier to work with data and design software for it. Many extant data sets such as TIMIT \citep{garofoloDARPATIMITAcousticphonetic1993} and Buckeye \citep{pittBuckeyeCorpusConversational2007} engage in this simplified practice for annotation, which makes it difficult to design computational systems that do anything else. An everpresent challenge exists for us as a field of making our file formats, software, and annotations reflect our knowledge of the nature of speech. We should rise to meet this challenge, and the method I detail here is one such step.

\subsection{The MAPS system}

A notable differentiating feature of MAPS is that it can provide boundary precision greater than the typical 10 ms limit. It provides the functionality to linearly interpolate between the probabilities around where the boundary would normally be inserted and place the boundary at the interpolated point instead. \citet{kelleyMasonAlbertaPhoneticSegmenter2024} noted that this technique greatly improved the performance of MAPS in terms of placing boundaries within 10 ms of their placement in the reference data.

MAPS was trained on the TIMIT \citep{garofoloDARPATIMITAcousticphonetic1993} and Buckeye speech corpora \citep{pittBuckeyeCorpusConversational2007}. The data were split into training, validation, and testing sets, and the amount of data, given in hours, is shown in Table~\ref{tab:time}. MAPS was trained 10 separate times with different initializations to be able to provide estimates of uncertainty regarding the model performance reported in \citet{kelleyMasonAlbertaPhoneticSegmenter2024}. In turn, those 10 models are being used in the present work as an ensemble for generating gradient boundaries through the construction of confidence intervals from order statistics.

\begin{table}[htb]
    \centering
    \caption{Breakdown of the amount of data, in hours, in the training, validation, and testing sets. The amount of time for both the TIMIT and Buckeye sets is given in addition to the total time.}
    \begin{tabular}{cccc}
    \toprule
    Set & \shortstack{TIMIT \\ time (h)} & \shortstack{Buckeye \\ time (h)} & \shortstack{Total \\ time (h)}\\
    \midrule
    Training & 3.75 & 15.32 & 19.07\\
    Validation & 0.18 & 1.25 & 1.43\\
    Testing & 1.44 & 1.94 & 3.38\\
    \bottomrule
    \end{tabular}
    \label{tab:time}
\end{table}

\subsection{The present paper}

In the remainder of the present paper, I first theoretically examine the nature of the boundaries that MAPS produces. Then, I motivate a nonparametric method for calculating a confidence interval around the median of the 10 boundaries the neural network ensemble emits for each segment, thereby producing gradient boundary regions. Finally, I present an empirical analysis of the gradient boundaries and their behavior for different categories of sounds.

\section{Theoretical analysis}

To properly estimate confidence intervals to use for the boundary regions, a theoretical backing for the phenomena being examined must be produced. The main reason for this is that there are multiple ways to derive confidence intervals based on choices about which statistical techniques to apply. Such choices might include whether to use parametric or nonparametric methods. The present section motivates the choice of a nonparametric method. Moreover, this section will make the nature of the uncertainty in the boundaries more explicit, which may help avoid misinterpretations over what the confidence interval itself for a boundary represents since it may not match intuition. The nature of boundaries from forced alignment is not straightforward to reason about since the time points for segment boundaries do not come directly from an acoustic model that outputs conditional probabilities of segment classes given some acoustic input. Rather, the time points come from subsequent processing of the model output.

It is important to note that the theoretical analysis presented here is specific to how MAPS performs forced-alignment with a neural network. The analysis may not apply entirely to systems that use a Gaussian mixture model with a hidden Markov model \citep[like, for example, the Montreal Forced Aligner,][]{mcauliffe_montreal_2017} or to transformer-based systems \citep[like  the Charsiu aligner,][]{zhuPhonetoaudioAlignmentText2022}.

\subsection{The nature of the boundaries MAPS produces}

For forced alignment systems that make use of boundaries from pre-aligned data \citep{gorman_prosodylab-aligner_2011,kelleyMasonAlbertaPhoneticSegmenter2024,mcauliffe_montreal_2017,kisler_signal_2012,yuan_speaker_2008}, a utilitarian assumption\footnote{I use the term ``assumption'' here in the same sense in which it is used for statistical modeling. Linear regression assumes linearity in the data, which is rarely true, but that does not stop linear regression from being a useful tool when analyzing data.} is made that these are ``true'' or ``gold standard'' boundaries, and the boundaries in the data used for training and evaluation are taken as such. This assumption is made whether the boundaries are used for training \citep[as in][]{kelleyMasonAlbertaPhoneticSegmenter2024} or just comparison during evaluation (as in most other aligners). Were this not the case, the evaluation strategies for forced alignment would lose meaning.

Yet, this assumption is clearly not veridical since there is no ``true'' point of demarcation between two segments. This point is trivially true because time is uncountably infinite, leading to a range of segmentations being acceptable, and more meaningfully so because some segment combinations like approximant-vowel, vowel-vowel, and vowel-approximant sequences inherently have no clear, inarguable moment of demarcation. Glottal stops also present similar difficulty in finding moments of demarcation since they are often expressed with a high degree of simultaneity with surrounding sounds \citep{garellekProductionPerceptionGlottal2013}. Still, this assumption is required in, again, a utilitarian and not realist sense to get off the ground. This is also not to say that the developers or users of forced alignment systems, nor the field at large, believe this to be true.

As a notational note in this section, following the use of Householder notation in \citet{vandegeijnAdvancedLinearAlgebra2020}, vectors, sets, and sequences are represented in lowercase Roman letters, and larger structures containing multiple vectors, sequences, etc. are represented in uppercase Roman letters. Scalars are represented in lower-case Greek letters that correspond to the Roman letters, e.g., $\chi$ is used for elements in a vector $x$, $\psi$ is used for elements in a vector $y$, and $\sigma$ is used for elements in a sequence $s$. Exceptions are made for traditional uses of Roman letters, where $P$ is used for a probability function, $n$ and $m$ are used for length/size, and $i$ and $j$ are used in reference to iteration or indexing.

The symbols will be defined as they are introduced, but they are also given here so they are all in one place. Concretely, the symbols have the following specific meaning in the context of the present system:

\begin{itemize}
    \item $P$ is a probability function.
    \item $\kappa$ is a single segment in the set of segments $k$ the acoustic model chooses between.
    \item $x$ is a vector of MFCCs, each forming part of the matrix $X$ that represents the MFCCs over time for an utterance.
    \item $\psi$ is a correct segment label for an MFCC vector $x$, and $y$ is the sequence of correct segment labels; these variables are primarily used in the training process for the acoustic model.
    \item $n$ is the length of $y$, or otherwise, the number of discrete time steps in an utterance.
    \item $\lambda$ is a label in the desired label sequence $l$.
    \begin{itemize}
        \item Note that the label sequence $l$ is different from $y$ in that $l$ does not assign the labels to each vector $x$ in $X$. So, $l$ might look like [k\ae{}t] for the word \textit{cat}, while $y$ might look like [kk\ae{}\ae{}\ae{}\ae{}tt].
        \item $\lambda$ and $l$ are primarily used in reference to alignment.
    \end{itemize}
    \item $m$ is the length of $l$.
    \item $s$ is a sequence of labels that collapses down to $l$, and $S$ is the set of all such sequences.
    \item $c$ is a sequence $s$ in $S$ with maximal probability.
    \begin{itemize}
        \item It is well known that there may be more than one sequence with maximal probability, so $c$ may not be a unique solution.
        \item $c$ can be viewed as a maximally probable approximation of $y$.
    \end{itemize}
    \item $c^\prime$ is a prefix of $c$, that is, the portion of $c$ that has been calculated up to a particular moment in time.
    \item $t$ is a vector of time points, each representing the end boundary for a particular phone.
\end{itemize}

Keeping in mind that true boundaries are being assumed, the acoustic model for MAPS is approximating a function producing a time-varying conditional distribution that outputs $P(\psi_i = \kappa \vert x_i)$ over all $\kappa \in k$, with $k$ as the set of segment classes the acoustic model has.\footnote{The acoustic model for other forced alignment systems are somewhat similar, but I focus exclusively on the model that is used in MAPS in the present paper.} Additionally, $\psi$ is the correct segment category and $i$ represents a discrete time step on the interval $[1,\, n]$ (with $n$ as the total number of time steps in an audio file). In prose, the function outputs a distribution of probability over all segment categories in the system, which are specified by the user during training. Note that, for recurrent neural networks (in contrast with fully-connected and convolutional neural networks), the way that $P$ is evaluated takes into account the sequential context of the moment where each individual $x$ occurs.\footnote{The traditional hidden Markov model+Gaussian mixture model combinations with triphones for forced alignment can account for some degree of sequential context.} In turn, the acoustic model is the backbone of a function that maps from the input data to the time points of the boundaries. This mapping is an approximating function for the relationship between the acoustic data, the segment labels, and the time points of the data used to train (i.e., acoustic data, segment labels) and evaluate (i.e., time points) the system.

When $P$ is evaluated over all $n$ time steps in an audio file, a matrix of probabilities is produced. The logs of the probabilities in this matrix are used in the alignment process to find a path that yields the maximal cumulative probability of labels that yields the specified label sequence $l$ (the transcription a user has specified). That is, through dynamic programming, the alignment algorithm \citep[such as the \texttt{Decode} algorithm in][]{kelleyMasonAlbertaPhoneticSegmenter2024} yields the solution $c$ defined as

\begin{equation}
c = \mathop{\arg \max}_{s \in S} \sum_{\sigma_i \in s} \log(P(\psi = \sigma_i \vert x_i))\,,
\end{equation}

\noindent where $S$ is the set of all sequences $s$ that map onto the length-$m$ label sequence $l$. That is, each $s$ will collapse into $l$ (the correct transcription) when duplicates are removed.

As a brief example, consider an audio file with five time-steps needing aligned labels, meaning that $n=5$ and all $s \in S$ are also therefore of length 5. Assume the correct transcription is [las] for Spanish \textit{las} `the.\textsc{fem.pl}', implying that $m=3$. There are several sequences of segments that would yield [las] after removing duplicates. These include \{laaas\}, \{llaas\}, and \{lasss\}. Each of those examples would be one instance of $s \in S$, and each symbol in each sequence would be a $\sigma_i \in s$. For instance, for \{laaas\}, $\sigma_1 = \text{l},\, \sigma_2=\text{a},\, \sigma_3=\text{a},\,\sigma_4=\text{a},\, \sigma_5=\text{s}$. If the sequence with the highest probability were \{laaas\}, we would have that $c=\{\text{laaas}\}$.

In sum, $c$ is the sequence of segment labels that collapses to the correct transcription with the highest overall probability given the acoustics. The solution $c$ is not necessarily unique, of course, and it is possible that there is more than one $s$ in $S$ that maximizes the probability.

Working from the sequence in $c$, a vector of time points $t$ of length $m$ is derived, where each $\tau \in t$ represents the latest moment for which

\begin{equation}
P(c^\prime * \lambda_j) > P(c^\prime * \lambda_{j+1})
\label{eq:concat}
\end{equation}

\noindent is true, constrained by $\tau_1 < \tau_2 < \ldots < \tau_m$. This probability is found by indexing into the dynamic programming matrix created during alignment. In this statement, $*$ is a concatenation operator, $c^\prime$ is the subsequence of $c$ before adding the next symbol $\sigma$ to the sequence, and $j < m$. Note that each $\lambda_j$ is an element in $l$, the correct transcription (or sequence of collapsed labels). Here, there is effectively a decision for whether to continue assigning the current segment symbol $\lambda_j$ or to proceed to the next one $\lambda_j+1$. The boundaries in the data being trained or evaluated on are taken to be each $\tau$ in $t$.

Returning to the example of Spanish [las], in (\ref{eq:concat}), it would be like choosing whether to continue believing the acoustic features are associated with [l] or proceed to say they are now associated with [a], which occurs at the latest moment that appending [l] to $c^\prime$ is more probable than adding [a]. Note that this example for \textit{las} is not equivalent to stating that a boundary occurs at the last moment that [l] is more probable than [a], which ignores the cumulative probability.

The last symbol at time point $n$ is a special case that is required to ensure that $\lambda_m$ occurs at least momentarily in the output. These statements are merely a symbolic reformulation of how the backtracing algorithm derives $c$ from the dynamic programming matrix.

Ultimately, each time point $\tau$ in the vector $t$ of boundary time points is a quantity that is being estimated using the conditional probabilities a system outputs.

To summarize, the boundaries in a forced alignment system represent the latest moment in time where it was more probable to append the current segment label to the cumulative sequence than the following segment label. Note that this is distinct from other possible segmentation schemes, like choosing the first moment when the following segment is more probable than the current one, which may result in premature boundaries since segment probabilities are not guaranteed to rise and fall smoothly throughout a particular utterance.

\subsection{Deriving gradient boundaries through confidence intervals}

A neural network learns an approximation of $P$, $\hat{P}$. Each time point derived from the alignment process is thus an approximation $\hat{\tau}$ for each $\tau \in t$. When several neural network approximations of $P$ are used together, multiple estimates are calculated for each $\tau$. Taken together, this ensemble can be treated as a sample of estimates of the time point $\tau$ produced using $P$, and a confidence interval can be derived, which can be used as a gradient boundary. I offer a forceful note that this confidence interval \textbf{should not} be interpreted as an interval around a ``true'' or ``correct'' boundary between two segments. Rather, the confidence interval is constructed for the approximating function between the input data and output time points. For some segment combinations, the optimal approximating function may not be able to place a reasonable boundary in some scenarios, and this is simply a shortcoming of forced alignment as a process.

In forced alignment, outliers are not uncommon, owing to, for example, acoustic models that do not match the data they are being used on well, poorly trained acoustic models, transcription errors, or simply a series of knock-on effects from one seemingly randomly misplaced boundary that the model must then make up for (much as a wave can move through traffic and build up from just one car suddenly braking). As such, a robust statistic like the median of the sample is more useful than the mean. As such, the boundary the ensemble produces should be placed at the median of $\hat{t}$.

From order statistics, using the 2nd and 9th ordered values (that is, the 2nd lowest and 9th highest values) as the extremes will produce an approximately 97.85\% confidence interval. Note that this confidence interval is not necessarily symmetric, unlike more familiar confidence intervals calculated based on the mean and standard error (which are more susceptible to outliers). It bears noting that that choosing the 2nd lowest and 9th highest values is strictly how this interval is constructed for a sample size of 10; if the sample size changes, such as by adding or removing models from the ensemble, the values to choose for a confidence interval of approximately 95\% will change. See \citet{hogg_probability_2015} for more information on constructing confidence intervals in this way, especially for other sample sizes.

Additionally, little is known about how best to train the approximation $\hat{P}$ since reasonable segment probabilities are difficult to derive \textit{a priori}. In practice, one-hot vectors are used for training, a choice which does not align with phonetic theory \citep{kelleyMasonAlbertaPhoneticSegmenter2024}. For this reason, one should note that the confidence intervals used for the boundary regions are based on approximations of distributions that are already less than ideal.

At the risk of being repetitious, it is important to keep in mind that these confidence intervals are estimates of uncertainty in the approximating function that produces time steps from the input data. That is, they represent how much the models in the ensemble vary in their estimate of the boundary. Care should be taken so as to not interpret this interval as a range derived through phonetic theory of where a segment boundary can be placed. The intervals are simply a representation of how certain or uncertain the model ensemble is in where the boundary goes, which permits representing boundaries as gradient regions. Subsequent analysis may find that there are useful links between the uncertainty and more theory-driven ranges for boundaries, but this assumption should not be made \textit{a priori}.

\section{Empirical analysis}

To test both changes in boundary placement and how well the uncertainty correlates with boundary error, an ensemble system was created using models that were trained during the development of MAPS. There were 10 models, and they were previously trained on a mix of data from the TIMIT \citep{garofoloDARPATIMITAcousticphonetic1993} and Buckeye \citep{pittBuckeyeCorpusConversational2007} speech corpora of American English, which have time-aligned phonetic transcriptions. The splits for training, validation, and testing follow the descriptions in \citet{kelleyMasonAlbertaPhoneticSegmenter2024}. The hidden layers in the neural networks were 3 long short-term memory (LSTM) layers with 128 units in each layer. The input was 13 MFCCs with the zeroth coefficient replaced by log energy and with delta and delta-delta features. The output was a set of American English segment categories. More details are given in \citet{kelleyMasonAlbertaPhoneticSegmenter2024}.

\subsection{Methods}

The general structure of generating the alignment for an utterance with confidence intervals proceeded as follows. First, each of the 10 previously trained models are used to create independent alignments for the utterance, using interpolation. Then, in the output representing the ensemble, for each boundary, the median from the 10 models' output for that boundary is selected as the point estimate of the boundary. As previously discussed, the 2nd lowest and 9th highest boundary values from the 10 models' output were then selected as the low and high ends of the boundary region, respectively. This process of selecting the median and interval points is repeated for each boundary for the utterance, and then the output files for the alignment are written to disk. During alignment, the interpolation technique from \citet{kelleyMasonAlbertaPhoneticSegmenter2024} was used on each model before ensembling to enhance the boundary placement accuracy and precision.

Alignment was made using the reference transcriptions given in the annotations for TIMIT and Buckeye, with the label foldings used in \citet{kelleyMasonAlbertaPhoneticSegmenter2024}, (e.g., folding all instances of TIMIT [k] closure, indicated \textlangle{}kcl\textrangle, into the [k] label). The error was calculated as both mean and median absolute error.

\subsection{Results}

The results are broken down into two main sections. The first is a more traditional error analysis that examines boundary error. The second explores the widths of the boundary regions that are being calculated.

\subsubsection{Error analysis}

I open this section by opining that global calculations like central tendencies of error as only so informative of how an aligner behaves. While they are traditional methods of evaluation, there are richer forms of analysis that can be performed, which will be given in the next section.

The boundary error results from the ensemble alignment are given in Table~\ref{tab:restab}. The adjustment given for the mean and median was to exclude the final boundary since it is placed at the end of the file and, thus, not informative of boundary error. Unadjusted values are retained for comparison purposes with \citet{kelleyBroadFineAcoustic2024}, and interpolation was used before the ensembling to improve the boundaries.

The overall values are generally a slight improvement over what \citeauthor{kelleyMasonAlbertaPhoneticSegmenter2024} reported. This improvement is a benefit of using the median of the ensemble since the median is robust to outliers. As a category, forced aligners can have extreme outliers from cascading errors based on maladaptive behaviors of the acoustic models, and the median ameliorates that problem to some degree.

\begin{table*}[!ht]
    \caption{Alignment boundary error results. Interpolation was performed before the ensemble process was applied. The system being used in the present paper is listed as ``M26'' in the ``System'' column, and it used the interpolation technique from \citet{kelleyMasonAlbertaPhoneticSegmenter2024}. The adjusted mean and median absolute error values were calculated after removing the final boundaries during error evaluation since the final boundaries are always placed at the end of the file and should always match the ``ground truth.'' The unadjusted metrics are retained for comparison with \citeauthor{kelleyMasonAlbertaPhoneticSegmenter2024}. The crisp interpolation results from \citeauthor{kelleyMasonAlbertaPhoneticSegmenter2024} are also provided for comparison at the end of the table, where the system is indicated as ``M24''.}
    \centering
    \begin{tabular}{cccccc}
        \toprule
        Data & System & \shortstack{Mean \\ abs. err. (ms)} & \shortstack{Median \\ abs. err. (ms)} & \shortstack{Adj. mean \\ abs. err. (ms}) & \shortstack{Adj. median \\ abs. err. (ms)} \\
        \midrule
        Train & M26 & 13.64 & 6.01 & 14.13 & 6.48 \\
        Val & M26 & 13.39 & 6.26 & 14.18 & 6.85 \\
        Test & M26 & 15.90 & 6.69 & 16.21 & 7.12 \\
        \midrule
        Train & M24 & $13.89 \pm 0.09$ & $6.36 \pm 0.04$ & --- & --- \\
        Val & M24 & $13.67 \pm 0.15$ & $6.65 \pm 0.05$ & --- & --- \\
        Test & M24 & $16.75 \pm 0.18$ & $7.14 \pm 0.005$ & --- & --- \\
        \bottomrule
    \end{tabular}
    \label{tab:restab}
\end{table*}

As additional error analysis, the breakdown of error by data set and corpus is given in Table~\ref{tab:breakdown}. The median test error for TIMIT is slightly lower than the values and outside the 95\% confidence interval in \citet{kelleyMasonAlbertaPhoneticSegmenter2024}; the Buckeye median absolute error is slightly higher and outside the confidence interval; and the combined median error is numerically higher but within the confidence interval. The mean absolute errors are all lower and all outside the reported confidence intervals. It is important to note two things about these results: 1) the improvements are minuscule even if potentially significant, and 2) having gradient boundary regions is the point of the proposed ensemble method, not an improvement in boundary placement. Nevertheless, the basic pattern of speaking style affecting alignment performance are still present in the data, where more casual speech as in Buckeye induces worse performance in alignment.

\begin{table*}[htb]
    \centering
    \caption{Boundary error breakdown by data set and corpus using the reference transcriptions and reference comparison method. The percentage of boundaries contained within the boundary region is also given. Related values from \citet{kelleyMasonAlbertaPhoneticSegmenter2024} are shown at the end of the table, marked with ``M24'', to facilitate comparison to previous work.}
    \begin{tabular}{ccccc}
        \toprule
        Set & Corpus & \shortstack{Mean abs. \\ err. (ms)} & \shortstack{Median abs. \\ err. (ms)} & \shortstack{Boundaries in \\ region (\%)} \\
        \midrule
        Train & TIMIT & 9.77 & 5.73 & 35.95 \\
        Train & Buckeye & 15.13 & 6.68 & 31.21 \\
        Train & All & 14.13 & 6.49 & 33.72 \\
        \midrule
        Val & TIMIT & 10.17 & 5.62 & 36.88 \\
        Val & Buckeye & 14.75 & 7.03 & 33.31 \\
        Val & All & 14.18 & 6.85 & 33.76 \\
        \midrule
        Test & TIMIT & 10.79 & 6.08 & 35.63 \\
        Test & Buckeye & 20.08 & 8.03 & 32.94 \\
        Test & All & 16.21 & 7.12 & 34.06 \\
        \midrule
        Test & M24 TIMIT & $11.51 \pm 0.16$ & $6.31 \pm 0.07$ & --- \\
        Test & M24 Buckeye & $20.47 \pm 0.27$ & $7.93 \pm 0.07$ & --- \\ 
        Test & M24 All & $16.75 \pm 0.18$ & $7.14 \pm 0.005$ & --- \\
        \bottomrule
    \end{tabular}
    \label{tab:breakdown}
\end{table*}

\subsubsection{Boundary region analysis}

\begin{figure}[htb]
    \centering
    \begin{subfigure}[t]{0.48\textwidth}
        \includegraphics[width=\textwidth]{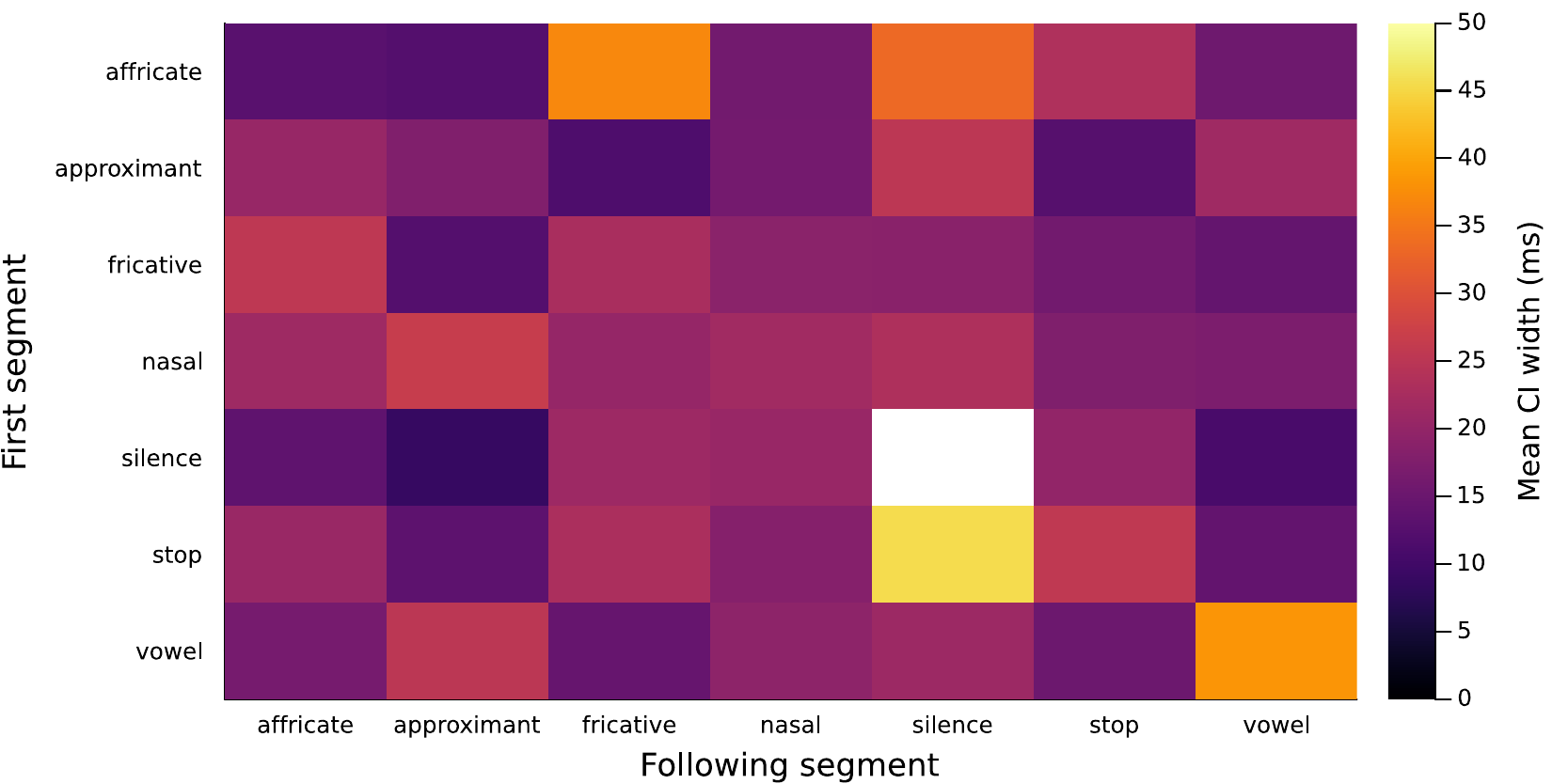}
        \caption{Mean boundary region width for the test set.}
        \label{fig:meanci}
    \end{subfigure}
    \begin{subfigure}[t]{0.48\textwidth}
        \includegraphics[width=\textwidth]{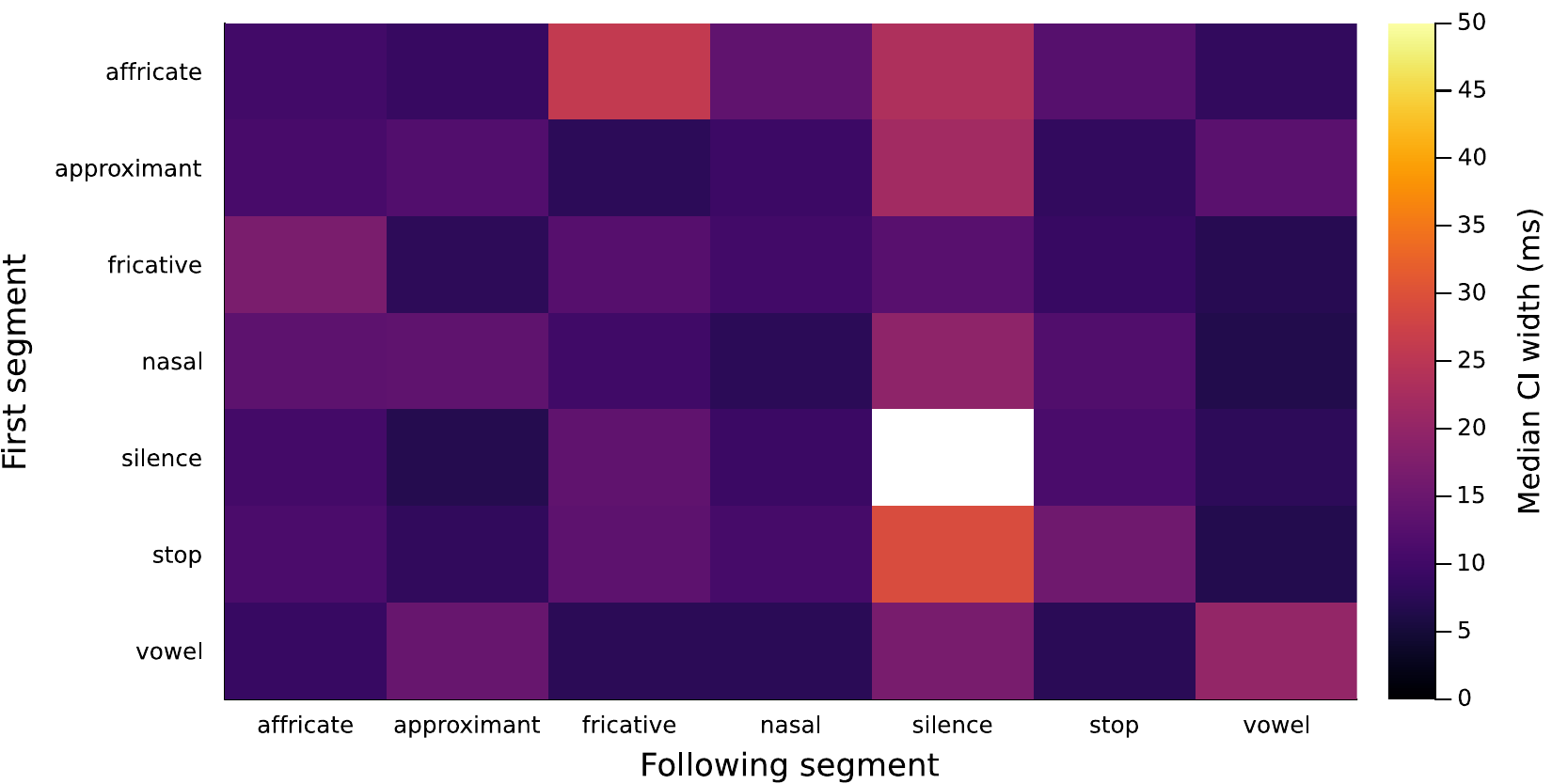}
        \caption{Median boundary region width for the test set.}
        \label{fig:mdnci}
    \end{subfigure}
    \caption{Heatmaps for central tendencies of the boundary region widths by bisegment type. The y-axes indicate the first category in a bisegment pair, while the x-axes indicate the second category in a bisegment pair. White cells indicate unobserved data; in this case, it is only silence-silence pairs that are unobserved, which the model cannot produce.}
    \label{fig:ciheat}
\end{figure}

To analyze the boundary regions, the widths of the confidence intervals were grouped by bisegment category pairs (e.g., a ``vowel-vowel'' bisegment sequence). Central tendencies were calculated for each and are shown in Figure~\ref{fig:ciheat}. Note that the range of values for the medians in Figure~\ref{fig:mdnci} is lower than that of the scale for the means in Figure~\ref{fig:meanci}, as observed in the overall cooler color of the plot. This pattern indicates again that the widths are sensitive to outliers. Moreover, the kinds of sequences that are typically assumed to be difficult to segment do indeed show greater widths, indicating greater uncertainty. This trend is particularly apparent in, for example, vowel-vowel and affricate-fricative sequences.

The proportion of boundaries contained in the boundary region in Table~\ref{tab:breakdown}, a metric \citet{zhangAmbulatoryMonitoringVocal2025} used, further underscore the previous theoretical discussion. Despite the boundary regions coming from confidence intervals, estimates of the limits have to do with the function the neural network was approximating, which is not strictly about placing boundaries. As such, researchers should interpret the confidence interval aspect of the regions with some caution; the estimated uncertainty relates to the system predictions, not the direct expression of a reasonable interval in which to place the boundary. The ensemble could be very certain and still place the boundary incorrectly.

Regarding the size of the boundary regions, they are somewhat sensitive to outliers in the data since it only takes three poor alignments across the ensemble to include an outlier as part of the region. This behavior can be seen in the difference between the mean and median width of the confidence intervals presented in Table~\ref{tab:cent}. The median width is noticeably smaller than the frame advance size of 10 ms, suggesting that the ensemble outputs were within one frame of each other more often than not.

\begin{table}[h]
    \centering
    \caption{Central tendencies for the size of the boundary regions.}
    \begin{tabular}{ccc}
        \toprule
        Data & \shortstack{Mean \\ width (ms)} & \shortstack{Median \\ width (ms)} \\
        \midrule
        Train & 14.09 & 7.53 \\
        Val & 14.95 &  7.94 \\
        Test & 17.07 & 8.54 \\
        \bottomrule
    \end{tabular}
    \label{tab:cent}
\end{table}

\begin{figure}[htb]
    \centering
    \includegraphics[width=0.8\columnwidth]{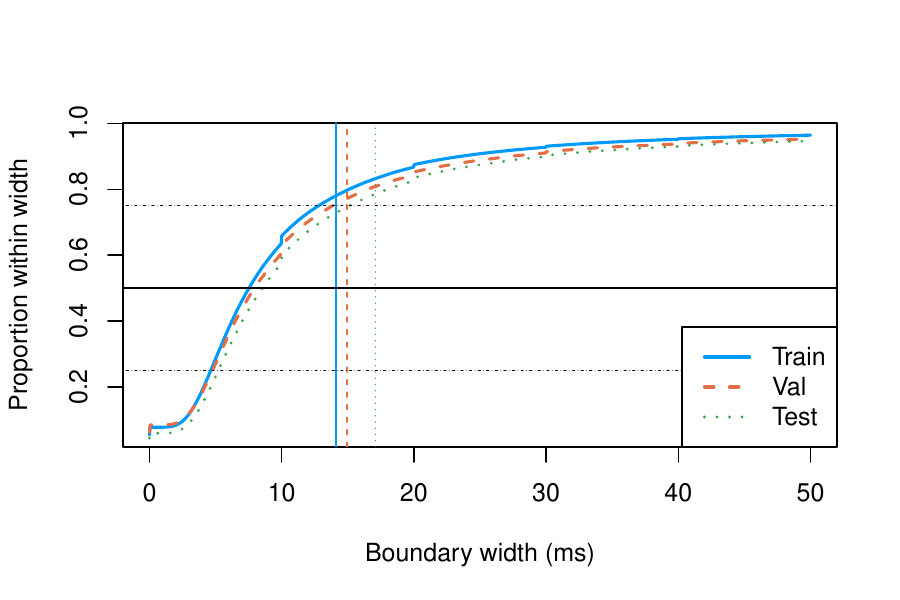}
    \caption{Empirical cumulative distribution function showing what proportion of the gradient boundary widths fall below a particular threshold. The training data is shown in blue with a solid line, the validation data in orange with a dashed line, and the the test data in green with a dotted line. The mean values for the widths of the sets are indicated with the vertical lines that have colors and line styles corresponding to the data set they represent. A black line is drawn at the 0.5 proportion, and the median widths for each set are where that line intersects the curves showing the distribution. The first and third quartiles are shown with a thin black line with a dash-dot pattern.}
    \label{fig:cdf}
\end{figure}

More detail on this pattern can be observed in Figure~\ref{fig:cdf}, where cumulative distribution functions (CDFs) show the proportion of widths that fall under a specific threshold. From visual inspection, there is a sharp rise in the proportion of data coverage between about 3 ms and 15 ms, after which the rise levels off. This trend indicates that the width of the bulk of the boundaries is in that range. Values outside this range likely tend toward indicating alignment issues.

\begin{figure*}[h!]
    \centering
    \begin{subfigure}[t]{0.48\textwidth}
        \includegraphics[width=\textwidth]{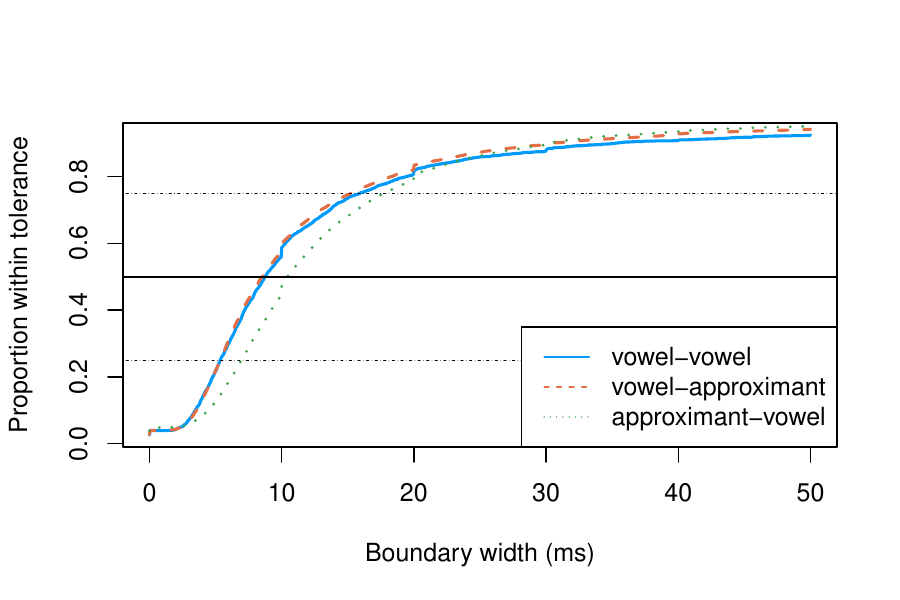}
        \caption{Comparison of vowel-approximant and approximant-vowel bigsegment categories to vowel-vowel.}
        \label{fig:vapp}
    \end{subfigure}
    \begin{subfigure}[t]{0.48\textwidth}
        \includegraphics[width=\textwidth]{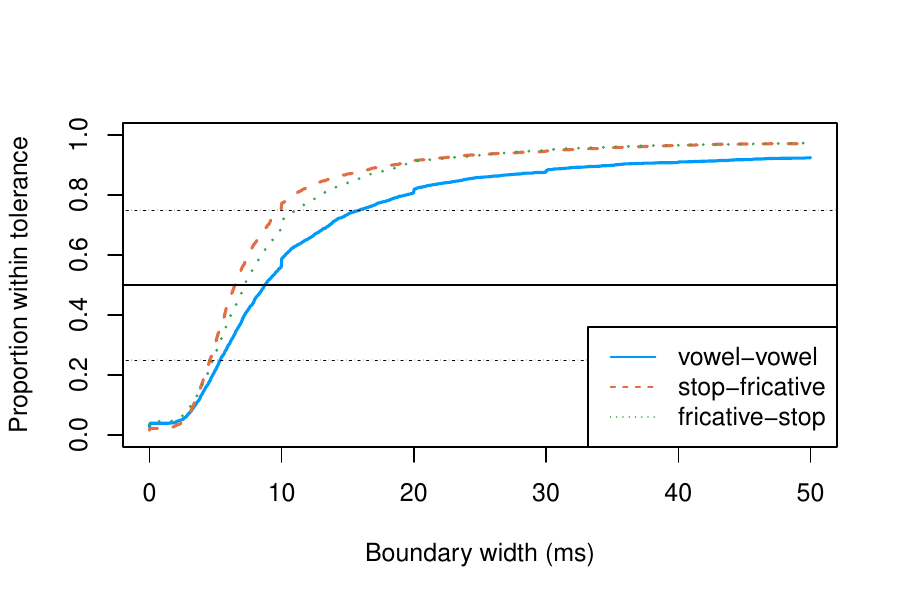}
        \caption{Comparison of stop-fricative and fricative-stop bigsegment categories to vowel-vowel.}
        \label{fig:vsf}
    \end{subfigure}
    \begin{subfigure}[t]{0.48\textwidth}
        \includegraphics[width=\textwidth]{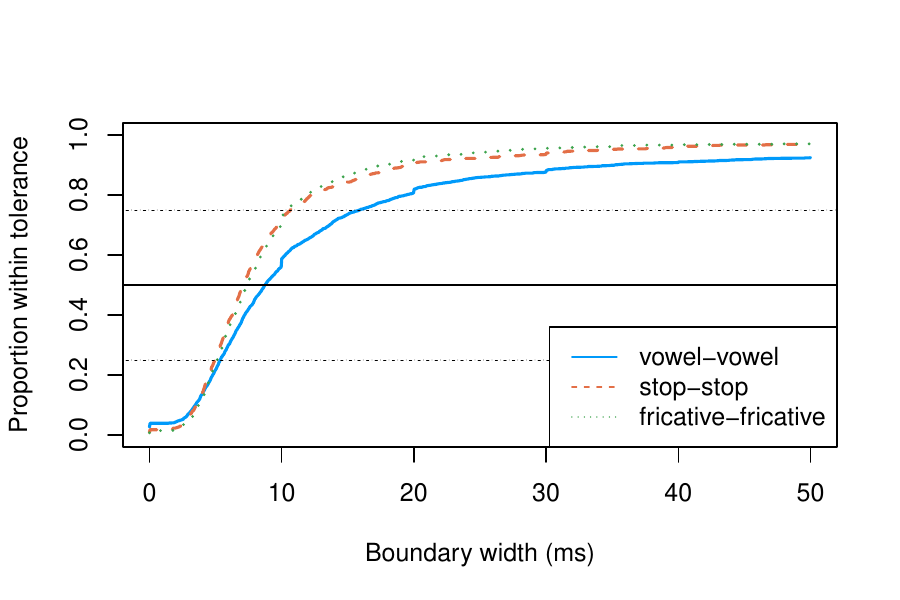}
        \caption{Comparison of stop-stop and fricative-fricative bigsegment categories to vowel-vowel.}
        \label{fig:vsf2}
    \end{subfigure}
    \begin{subfigure}[t]{0.48\textwidth}
        \includegraphics[width=\textwidth]{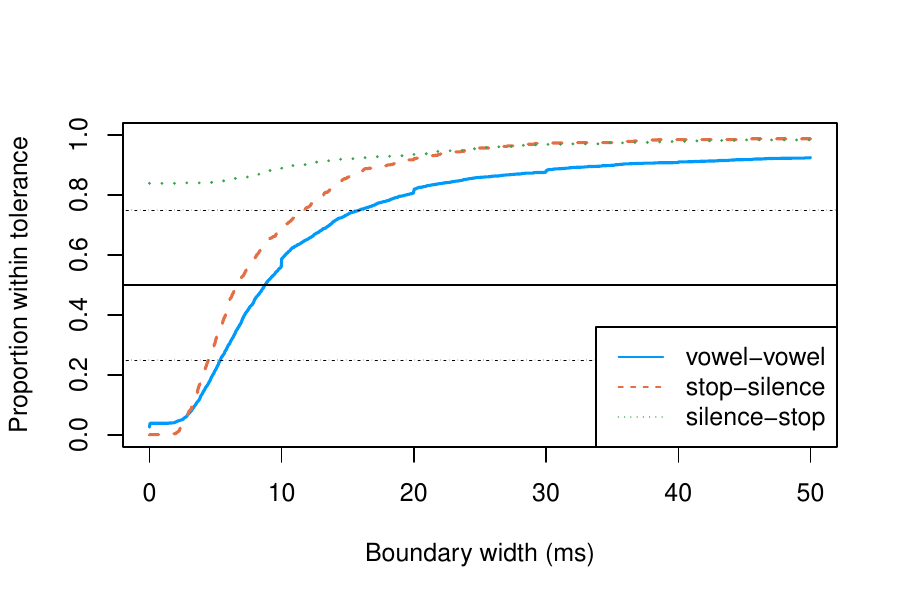}
        \caption{Comparison of stop-silence and silence-stop bigsegment categories to vowel-vowel.}
        \label{fig:vss}
    \end{subfigure}
    \caption{Empirical cumulative distribution function showing what proportion of the gradient boundary widths fall below a particular threshold for selected bisegment categories for the test data. The vowel-vowel bisegment results are retained throughout each plot to anchor the results. A line under the vowel-vowel line indicates a boundary region containing more ambiguity than vowel-vowel segmentation. Conversely, a line above the vowel-vowel line indicates a boundary region containing less ambiguity than vowel-vowel segmentation. Line color and type indicate different categories as given in the legend. A black line is drawn at the 0.5 proportion, and the median widths for each set are where that line intersects the curves showing the distribution. The first and third quartiles are shown with a thin black line with a dash-dot pattern.}
    \label{fig:typecdf}
\end{figure*}

In Figure~\ref{fig:typecdf}, a selection of more fine-grained results can be seen from the test data. The empirical CDF functions the distribution of the widths of the boundaries for different bisegment combinations. In Figure~\ref{fig:vapp}, it is striking that the approximant-vowel sequences overall show more ambiguity than vowel-vowel and approximant-vowel sequences through consistently wider boundary regions. It is possible that this is due to tautosyllabic vowel-glide sequences being impermissible in the data\footnote{In the transcription scheme used in TIMIT and Buckeye, diphthongs are treated as a segment category in their own right and not as vowel-glide sequences}, making vowel-liquid sequences more common (and possibly being less ambiguous than vowel-glide sequences). However, it may also be the case that there is something inherently more ambiguous or gradient in productions of approximant-vowel sequences than in vowel-vowel and vowel-approximant sequences.

From Figures~\ref{fig:vsf}, \ref{fig:vsf2}, and \ref{fig:vss}, stop-fricative, fricative-stop, stop-stop, fricative-fricative, stop-silence, and silence-stop sequences are all less ambiguous that vowel-vowel sequences, which matches phonetic intuition. Of note, however, is that the silence-stop sequence is markedly less ambiguous than stop-silence sequences. I believe that this is due to transcription conventions not generally including silence before an utterance initial stop as part of the stop, so it is relatively easy to segment that combination since the data was extracted to contain only single utterances, improving the overall behavior of the system for that combination.

\begin{figure*}[htb]
    \centering
    \begin{subfigure}[t][][t]{0.48\textwidth}
        \includegraphics[width=\textwidth]{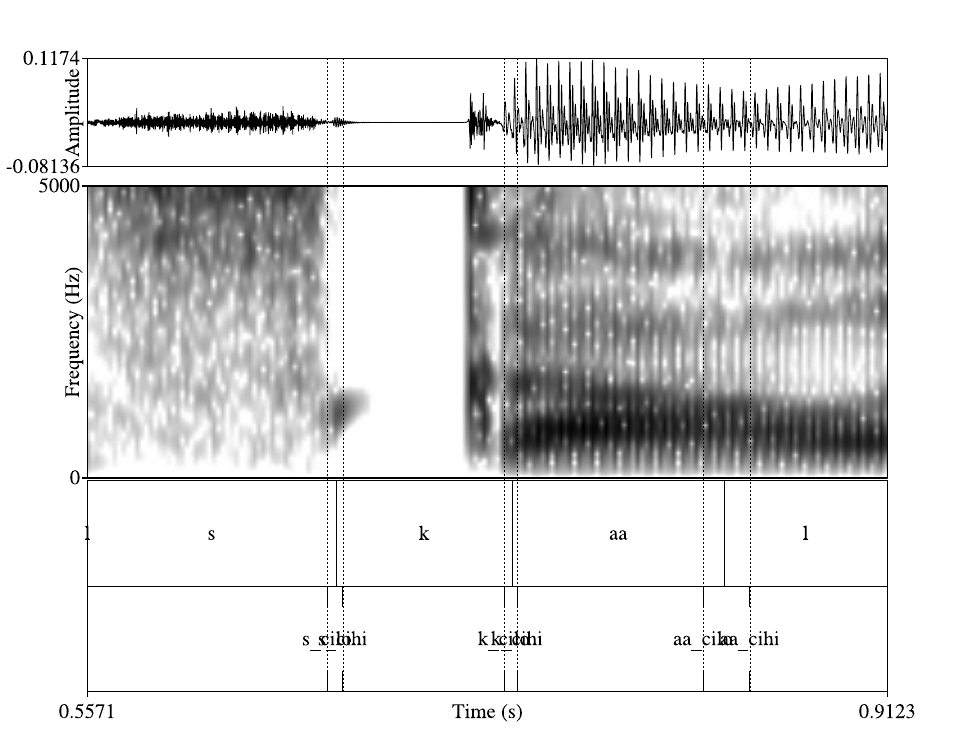}
        \captionsetup{width=0.8\linewidth}
        \caption{Offset boundaries of segments of \textup{[sk\textipa{A}]} from \textup{scholars} from ``biblical scholars argue'' in the TIMIT corpus. The natural overlapping of the labels in the TextGrid is shown.}
        \label{fig:wordseg}
    \end{subfigure}
    \begin{subfigure}[t][][t]{0.48\textwidth}
        \includegraphics[width=\textwidth]{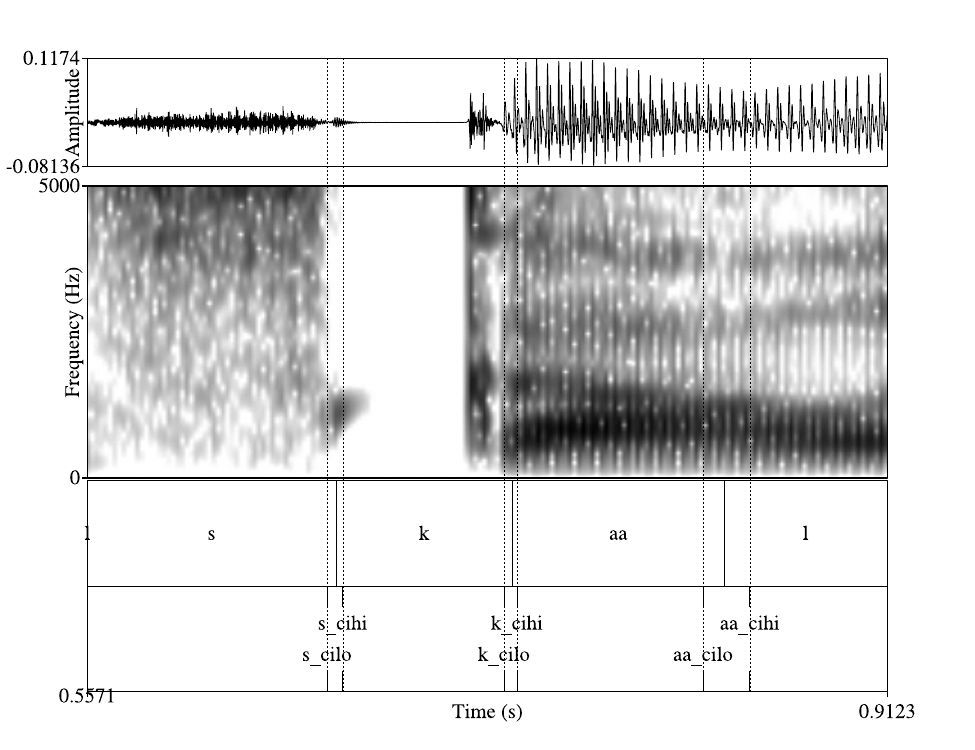}
        \captionsetup{width=0.8\linewidth}
        \caption{Offset boundaries of segments of \textup{[sk\textipa{A}]} from \textup{scholars} from ``biblical scholars argue'' in the TIMIT corpus. The labels were edited from what Praat produces to be offset to better show the labels for the boundary region limits.}
        \label{fig:segci}
    \end{subfigure}
    \caption{Sample of TextGrid format for segmentation of the first three segments in \textup{scholars} from sentence ``SX42'' from speaker ``FAEM0'' in the TIMIT corpus. The dashed lines indicate the limits of the boundary regions.}
    \label{fig:alignment}
\end{figure*}

Finally, a TextGrid sample of the alignment is given in Figure~\ref{fig:alignment}. The boundaries in the figure are largely agreeable. The confidence intervals in Figure~\ref{fig:wordseg} can be hard to see since they are narrow, though this is an inherent problem to using TextGrids for visualizing tiers that are on different time scales. An offset version is given in Figure~\ref{fig:segci} to better show the annotations.

Note how the intuition of [\textipa{A}l] in \textit{scholars} being harder to segment or the boundary being more ambiguous than [sk] is captured in the widths of the boundary regions. This is to say that the behavior of the ensemble of aligners produces results familiar to speech scientists and suggests that the acoustic models have learned patterns related to how acoustic patterns are expressed and conceptualized in phonetic theory.

\section{Discussion}

The use of the ensemble of networks is an improvement over using a single network. This is in line with neural network theory and previous findings \citep{breimanBaggingPredictors1996,goodfellowDeepLearning2016}. The gained resilience to outlying boundaries should help researchers get more consistent segmentation. Of course, this resilience is intertwined with the quality of the acoustic model and the transcription. Using the median of the ensemble boundary estimates does not categorically prevent cascading errors from happening, nor does it completely prevent a poor alignment. Regardless, it should still yield improved boundary placements over the singleton model.

The boundary regions derived from the ensemble are a useful tool for phonetic research. The width of the regions can be a useful heuristic for researchers to detect errorful alignments for manual correction. Moreover, it may be possible to use the uncertainty expressed in the confidence interval width to identify transcriptions where there may be more disagreement due to inherent uncertainty, such as over sonorant-sonorant or glottal stop boundaries, much as Figure~\ref{fig:ciheat} suggested. It may also prove useful to see how the ensemble of models compares with an ensemble of human annotators.

The boundary regions from confidence intervals are also generally more useful than the log likelihoods that can be output from the alignment process. Confidence intervals have absolute interpretations and meaningful units (here, seconds or milliseconds), while log likelihood values require a comparison to be meaningfully interpreted. Moreover, the boundary regions as confidence intervals are a direct statistical expression of model uncertainty, while log likelihood values are not. That is, a confidence interval can be interpreted to mean that 97.85\%\footnote{Note that this value is specific to using order statistics with a sample size of 10.} of the time the interval is constructed, the true value of the median will be contained in it.\footnote{Keeping in mind that the median here is related to the function the neural network is approximating, not necessarily an indication of the true boundary.} No such interpretation of the log likelihood is possible.

With that said, I want to re-emphasize that the point of the process described in the present paper is to view segment transitions as gradient, expressed as boundary regions calculated through confidence interval estimation. The main benefit of performing this process is greater correspondence between data annotations and the field's understanding of speech. There is also a richer possible statistical interpretation of the results of alignment and potentially comparing alignments between systems, much in the same way that having confidence intervals in statistical plots adds information in the groups being compared. Having boundary regions calculated through confidence intervals is also step down a particular route toward being able to tell if boundaries placed by different systems are statistically significantly different from each other. In addition, it is simply responsible science to attempt to estimate uncertainty in measurements, which in this case reflect a sense of ``model certainty'' in the alignment.

Theoretically, having gradient boundary regions accords better with the field's understanding of speech than do discrete points. The region that is highlighted by the boundary region relates well to indicating the transitions between segments and how that transition process is not discrete. I think again back to Figure~\ref{fig:ciheat} and Figure~\ref{fig:typecdf} where the aligner ensemble showed less agreement (indicating by wider boundary regions) on the point where one vowel stopped and became another. Moving from one vowel to another in speech is clearly a gradient process, and our data formats should match that observation. The ability to create these kinds of regions is crucial to improving the match between the tooling we use to study speech and our general understanding of it. As a field, we should strongly consider what forms of software and data formats afford the functions we need to analyze data. There is a risk that forms of annotation that do not have high correspondence with theory in speech impose limits on the research that is performed. One example might simply have to do with when research is met with the practicalities of the living world, where a particular research question may not end up investigated because it is difficult to invent a necessary annotation style that plugs in well with extant software, causing other research questions to be selected due to less overall work burden.

One drawback of using the ensemble approach is that the alignment must be performed several times---in this case, ten. For the present case, it produces approximately a tenfold increase in the time it takes to perform the alignment. While multiprocessing could, in theory, alleviate this problem since the ensemble alignment is easy to parallelize, having the requisite hardware---whether locally or in the cloud--to do this may be too costly to be viable for many researchers in speech science. Future work should explore speedups to increase the feasibility of calculating confidence intervals. With that being said, the alignment process happens faster than real-time, so it should be practical to align modest data sets with this technique.

Future research should explore other techniques by which estimates of uncertainty for alignments could be derived. One such example might be using Bayesian neural networks, which can yield estimates of uncertainty through credible intervals. However, the way in which the uncertainty in the network's predictions plays into temporal aspects of the segmentation is not as straightforward. It would also behoove the research community to have a segmentation format that more naturally allows for confidence intervals (or similar), whether through an extension to Praat TextGrids or a new format altogether. It may also be useful to see how boundary regions or measures of ambiguity or uncertainty can contribute to quantifying the rate of cascading or catastrophic alignment errors.

An anonymous reviewer pointed out that a feature related to confidence intervals that may also prove useful would be to dynamically change the alignment resolution after placing the initial boundary. What is derived in this way could also be treated as a region of plausible boundary placement based on the model output. Along these lines, other techniques for determining gradient boundary regions that do not involve statistical calculations of model certainty may also prove useful, especially if derived from phonetic theory and interpretable as phonetically plausible regions.

One point that should be emphasized is that Praat TextGrids are not an ideal format for representing this kind of confidence interval information. Segmentation in TextGrids is typically stored in an interval tier, which, by the way they are defined, can only represent a series of non-overlapping events. The events may be given a label, a start time, and an end time. There is no capacity for adding information about uncertainty of the start and end times, for example. Point tiers, by contrast, cannot represent temporal extents of events, and they cannot be explicitly connected to intervals in another tier. Regardless, TextGrids are output by default since Praat is a common touchpoint for the speech research community, and it is very useful for creating visualizations, though it seems unwise for TextGrids to be the only output file format.

Following on from this line of thought, the alignment system also outputs two other data formats. The first is a table with all segments and boundary region limits from a particular run, which can be used for statistical analysis. The second format is a JSON for each aligned file that provides similar information as a TextGrid but lifts the requirement that the events represented in the file format are non-overlapping. The JSON format is intended to be more useful for programmatic analysis than the TextGrid, owing to the emitted TextGrid format being incapable of tightly pairing the confidence interval information with the segmentation boundaries. Both the table and JSON represent a view of speech events where the temporal extent one speech event is not limited by the temporal extent of other speech events, which that is less constrained than view one suggested by the interval and point tiers in Praat TextGrids. The entries in the table and JSON simply represent a set of speech events, each with their own collections of information.

I note and recognize that the use of TextGrids by the speech research community does not suggest that the field necessarily endorses a non-overlapping view of speech, and it is difficult to imagine how overlapping events might be meaningfully displayed as data annotations. Still, it would benefit the field to think more explicitly about what mathematical and data structure(s) might be most appropriate to contain annotations about speech events \citep[similar to some ideas expressed by][Chapter 2, in examining the relation between transcriptions and temporal precedence and overlap]{coleman1998phonological}. Making choices along these dimensions constrains what is possible to represent, and the field should be careful not to tie its hands by adhering too faithfully to unwarranted file format limitations. Less constrained file formats might not prove to be an obstacle since physical constraints on speech production still come to bear on phonetic data.

Code used to perform the data preparation and analyses in the present paper is available on GitHub \url{https://github.com/maetshju/maps_ci_analysis} archived through Zenodo \url{https://doi.org/10.5281/zenodo.15571293}. The confidence interval features are also available in current releases of MAPS.

\section{Conclusion}

Speech segmentation is a naturally gradient and uncertain process. Neither previous forced alignment systems nor current annotation conventions have represented gradience in their outputs. The ensemble modification to MAPS and transcription scheme allow researchers to estimate and make use of gradient boundary regions based on confidence intervals on the boundaries. Having regions of any kind associated with segment boundaries also represents a step toward more gradient or overlapping segmentation schema, which would be more in line with representational frameworks where overlap is permitted in articulatory \citep{browmanArticulatoryPhonologyOverview1992} or acoustic \citep{kelleyBroadFineAcoustic2024} dimensions.

\section*{Acknowledgments}

I would like to thank members of the Mason Phonetics and Phonology Lab for discussion of previous versions of this topic, as well as the attendees of the 188th Meeting of the Acoustical Society of America for their feedback on this project. I would also like to extend thanks to Pertti Palo for identifying bugs and shortcomings in the user-interface for this feature in MAPS. I am additionally grateful to two anonymous reviewers and the associate editor for their insightful comments and genuine engagement with the paper. All remaining errors are, of course, my own.

\section*{Ethics statement}

No experiments with human or animal subjects were performed
as a part of this study, and no ethics approval was needed for the present study.

\section*{Conflict of interest statement}

The author has no conflicts of interest to report.

\section*{Author contributions}

Matthew C. Kelley: conceptualization, methodology, software,
validation, formal analysis, investigation, data curation, writing – original draft,
writing – review and editing, visualization, project administration

\printbibliography

\end{document}